\documentclass[preprint,twocolumn]{jpsj2}

\usepackage{amsmath}
\title{%
Aharonov-Bohm Effect for Parallel and  T-shaped
Double Quantum Dots
}

\author{%
Yoichi {\sc Tanaka}  and  Norio {\sc Kawakami}
}

\inst{%
Department of Applied Physics, Osaka University, Suita, Osaka 565-0871
}

\recdate{\today}

\abst{%
We investigate the Aharonov-Bohm (AB) effect for the double quantum dots 
in the Kondo regime using the slave-boson mean-field approximation.
In contrast to the non-interacting case,
where the AB oscillation  generally has the period of 4$\pi$
when the two-subring structure is formed via the interdot tunneling $t_c$,
we find that the AB oscillation  has the period of 2$\pi$ in the Kondo regime.
Such effects appear for the double quantum dots close to
the T-shaped geometry even in the charge-fluctuation regime. 
These results follow from the fact that the Kondo resonance
is always fixed to the Fermi level irrespective of the detailed structure
of the bare dot-levels.  
}

\kword{%
Aharonov-Bohm effect, Kondo effect, Double Quantum Dots
}

\begin{document}
\maketitle


Recent rapid progress in the nanotechnology makes it 
 possible to fabricate double quantum dot (DQD) systems.
\cite{Wiel:2003}
One of the remarkable features in the DQD systems is the interference effect
induced via multiple paths of electron propagation,
which is very sensitive to the external magnetic 
flux, giving rise to the Aharonov-Bohm (AB) effect
\cite{Holl:2001,Sigr:2004}.
The AB effect has been theoretically studied  for the DQD
with\cite{Akera:1993,Izumida:1997,Boese:2002,Utsumi:2004,Lopez:2005}
 or
without\cite{Kubala:2002,Jiang:2002,Orellana:2004,Kang:2004,Zhi:2004}
the intradot Coulomb interaction.
As noted in the literature 
recently,\cite{Jiang:2002,Orellana:2004,Kang:2004,Zhi:2004}
the interdot tunneling between two dots, which forms two-subring
 structure, generates a new AB oscillation period of $4\pi$
instead of the normal period of $2\pi$.
However, the arguments are based on the non-interacting electron
model, and  the influence of electron correlations
on this new AB oscillation has not been discussed yet.

\begin{figure}[h]
\begin{center}
\includegraphics[scale=0.3]{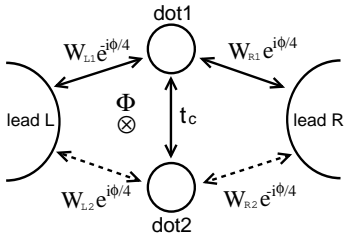}
\end{center} 
\caption{
DQD system with the magnetic flux $\Phi$.
}
\label{f1}
\end{figure}

In this note, we address this question about the AB 
effect by explicitly taking into account the 
Kondo effect in the DQD shown in Fig. \ref{f1}. 
\vspace{-3mm}
\vspace{-2mm}
We will show that if the Kondo effect is taken into account
in the parallel DQD, the AB
period should be practically regarded as $2\pi$ instead of $4 \pi$.
This tendency becomes more remarkable in the DQD close to the
 T-shaped geometry, which is another typical 
DQD system sensitive to the interference effect.

In the following discussions, the intradot Coulomb
interaction is assumed to be sufficiently large,
so that double occupancy at each QD is forbidden. 
This assumption 
which allows us to use a slave-boson
representation of correlated electrons in the dots.
In this representation, 
the DQD system in Fig. \ref{f1} is modeled
with the $N$(=2) fold 
degenerate Anderson Hamiltonian 
\begin{eqnarray}
H&=&\sum_{k_\alpha ,\sigma}\varepsilon _{k_\alpha}
c_{k_\alpha\sigma}^\dag c_{k_\alpha\sigma}^{}
 +\sum_{m,\sigma}\varepsilon _{d}
f_{m\sigma}^\dag f_{m\sigma}^{}
\nonumber\\
&+&\frac{t_c}{N}\sum_{\sigma}(f_{1\sigma}^\dag b_1^{}
b_2^\dag f_{2\sigma}^{}
+h.c)
\nonumber\\
&+&\frac{1}{\sqrt{N}}\sum_{k_\alpha,m,\sigma} (V_{\alpha m}
c_{k_\alpha\sigma}^\dag b_m^\dag f_{m\sigma}^{}+ h.c.)
\label{Hami}
\end{eqnarray}
where $c_{k_\alpha\sigma}$ is the annihilation operator
of an electron with spin $\sigma$
in the lead $\alpha$ ($\alpha=L, R$),
and the annihilation
operator of an electron in the dot $m$ ($m=1,2$)
with the energy $\varepsilon_d$
is represented as $ b_{m}^\dag f_{m\sigma}^{} $
 with the constraint,
$f_{m\sigma}^\dag f_{m\sigma}^{}+b_m^\dag b_m^{} =1$,
where $b_{m}$ ($f_{m\sigma}$) is the slave-boson
(pseudo-fermion) annihilation operator for an empty state
(singly occupied state).
 We use the mean-field approximation at zero temperature
(see ref.\citen{tanaka:2005}).
The effect of the magnetic flux $\Phi$ is symmetrically incorporated in
the tunneling between the dot $m$ and the lead $\alpha$ in eq. \eqref{Hami}, 
$V_{L1(R2)}=W_{L1(R2)}e^{-i\phi /4}$,
$V_{L2(R1)}=W_{L2(R1)}e^{i\phi /4}$,
where $\phi =2\pi\Phi/\Phi_0$ ($\Phi_0=h/e$).
This means that the magnetic flux equally pierces the two subrings formed by the interdot tunneling $t_c$. 
We systematically change the arrangement of the DQD by modifying the resonance width $\Gamma_{1,1(2,2)}^\alpha =\pi |W_{\alpha 1(2)}|^2 \rho$ ($\rho$ is the density of states for leads at the Fermi energy): 
the parallel DQD is given by 
$\Gamma_{1,1}^{\alpha} \sim  \Gamma_{2,2}^{\alpha}$, and
the T-shaped DQD 
 by $\Gamma_{2,2}^{\alpha} \sim 0$ with fixed 
 $\Gamma_{1,1}^{\alpha}$.\cite{Taka}

\begin{figure}[h]
\begin{center}
\includegraphics[scale=0.3]{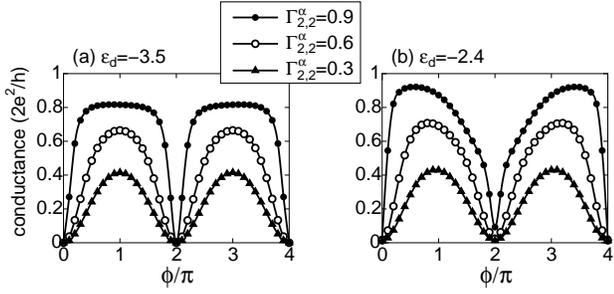}
\end{center} 
\caption{Conductance as a function of the magnetic flux $\phi$: 
(a) Kondo regime, (b) charge-fluctuation 
regime. We choose the interdot coupling $t_c=1.5$
in the unit of $\Gamma_{1,1}^{\alpha}$ (=1).}
\label{f2}
\end{figure}
Figure \ref{f2} shows the conductance as a function of the magnetic 
flux $\phi$ for several values of $\Gamma_{2,2}^{\alpha}$.
We have studied two typical cases in the Kondo regime and 
the charge-fluctuation regime. 
There are several remarkable properties in the conductance. 
(i) As seen in Fig. \ref{f2}(a), the AB period of the conductance 
seems to be always $2\pi$ in the Kondo regime ($\varepsilon_d=-3.5$) 
even if we change the arrangement of the  DQD from the parallel to
T-shaped geometry. This is in contrast to the conclusion for 
the non-interacting
case: the period should be  $4\pi$ generally,
\cite{Jiang:2002,Orellana:2004,Kang:2004,Zhi:2004} which
follows from the two-subring structure formed 
via the interdot tunneling $t_c$.
(ii) On the other hand, in the charge-fluctuation 
regime ($\varepsilon_d=-2.4$) in Fig. \ref{f2}(b),
the AB oscillation for the parallel DQD 
($\Gamma_{2,2}^{\alpha}=0.9$) indeed has the period of 
$4\pi$. Even in this case, however, as the DQD system
is changed from the  parallel to T-shaped geometry,
the period again approaches $2\pi$
(see $\Gamma_{2,2}^{\alpha}=0.3$ in Fig. \ref{f2}(b)).
These results imply that electron correlations play a
crucial role for the AB oscillation in the DQD.

\begin{figure}[h]
\begin{center}
\includegraphics[scale=0.3]{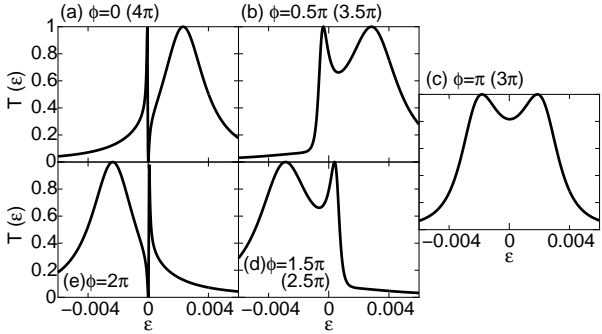}
\end{center} 
\caption{Transmission probability $T(\varepsilon)$ 
around the Fermi energy ($\varepsilon=0$)
 in the Kondo regime ($\varepsilon_d=-3.5$).
We set $t_c=1.5$ and $\Gamma_{2,2}^{\alpha}=0.9$.}
\label{f3}
\end{figure}
The above characteristic properties result from
 the Kondo resonances affected by the interference  effect.
In Fig. \ref{f3} the transmission probability $T(\varepsilon)$
 is shown around the Fermi energy ($\varepsilon=0$)
in  the case close to the parallel geometry
($\Gamma_{2,2}^{\alpha}=0.9$). At $\phi=0$, 
the Kondo resonance splits 
into two distinct resonances due to the interdot coupling $t_c$:
 they consist of a sharp bonding Kondo resonance at the Fermi energy
and a broad anti-bonding Kondo resonance\cite{tanaka:2005}.
The interference between these two resonances 
produces a dip in $T(\varepsilon)$ around $\varepsilon=0$, as 
seen in Fig. \ref{f3}(a).
As the magnetic flux $\phi$ increases, the interference effect is 
smeared (Fig. \ref{f3}(c)), and
the Kondo resonances in $T(\varepsilon)$
become almost symmetric at $\phi=\pi$. 
At $\phi=2\pi$, the shape of
$T(\varepsilon)$ is inverted with respect to the Fermi energy
as shown in Fig. \ref{f3}(e), and then
$T(\varepsilon)$ returns to the original shape
with another increase by $2\pi$. We thus see
that $T(\varepsilon)$ changes with the period of $4\pi$,
as should be expected. However, it should be noted that 
the value of $T(\varepsilon=0)$ oscillates with the period 
very close to $2\pi$,
which characterizes the linear conductance 
shown in Fig. \ref{f2}(a) for $\Gamma_{2,2}^{\alpha}=0.9$.
This new periodicity is due to the constraint imposed on the
 Kondo resonances: the self-energy shift inherent in the Kondo
effect properly tunes the position of the resonances so as to keep
the electron number in the dots constant.
This many-body effect gives rise to the effective period of $2\pi$,
in contrast to the non-interacting case.
\cite{Jiang:2002,Orellana:2004,Kang:2004,Zhi:2004}


\begin{figure}[h]
\begin{center}
\includegraphics[scale=0.3]{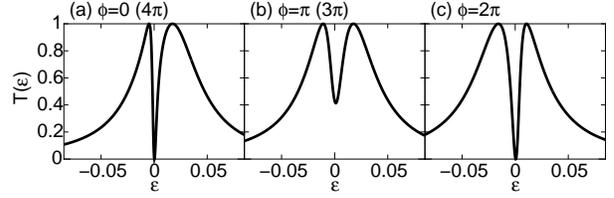}
\end{center} 
\caption{Transmission Probability $T(\varepsilon)$ for 
$\Gamma_{2,2}^{\alpha}=0.3$ in the charge-fluctuation regime
($\varepsilon_d=-2.4$). 
}
\label{f4}
\end{figure}
\vspace{-3mm}
In the charge-fluctuation regime ($\varepsilon_d=-2.4$),
the resonances are both lifted slightly above the Fermi energy.
The resulting asymmetric nature in $T(\varepsilon)$ naturally gives
the AB effect with the
 period of $4\pi$, as seen 
for $\Gamma_{2,2}^{\alpha}=0.9$ in Fig. \ref{f2}(b).
However, even in this charge-fluctuation regime, we encounter a remarkable 
fact  that when the system approaches the T-shaped 
DQD ($\Gamma_{2,2}^{\alpha}=0.3$ in Fig. \ref{f2}(b)), 
the effective period becomes $2 \pi$ again.
To understand this tendency, we
should recall that while the whole DQD system is in the 
charge-fluctuation regime,
the dot-2 itself falls into the Kondo regime 
because it is almost decoupled from the leads in the T-shaped geometry. 
We can see that
the broad resonance of the dot-1
 develops a dip structure at the Fermi energy, which is caused by
 the sharp Kondo resonance formed in the dot-2.
In this case, the DOS of the dot-1 mainly controls the 
transmission probability $T(\varepsilon)$ shown in Fig. \ref{f4}.
When the magnetic flux $\phi$ is turned on, the dip in $T(\varepsilon)$ 
is slightly modified with the resonance-width almost unchanged.
Therefore, although the current flows
 mainly through  the dot-1 in the T-shaped geometry,
  the sharp dip structure
caused by the Kondo effect in the dot-2 essentially
controls the linear conductance in the system. This gives rise to 
 $2\pi$ for the effective AB period
even in  the charge-fluctuation regime,
as observed in Fig. \ref{f2}(b).


Finally, we would like to comment on the AB period in more general cases for which the magnetic flux piercing the two subrings is not equal: the ratio of flux is assumed to be an irreducible fraction $m/n$. In the non-interacting case, the AB period of the conductance is $2\pi(m+n)$. In the Kondo regime, however, the AB period becomes half, $\pi(m+n)$, provided that both $m$ and $n$ are odd, while it remains $2\pi(m+n)$ if either $m$ or $n$ is even.

In summary, we have shown that in spite of the two-subring 
structure in our DQD system, 
the AB oscillation practically has the period of $2\pi$ in the Kondo regime.
Such effects appear for the DQD close to
the T-shaped geometry even in the charge-fluctuation regime. 
This conclusion, which is contrasted to the noninteracting case,
 \cite{Jiang:2002,Orellana:2004,Kang:2004,Zhi:2004}
arises from the self-energy shift
inherent in the Kondo effect: the Kondo resonance is always fixed 
to the Fermi level irrespective of the detailed structure of
 the bare dot-levels.



\end{document}